\RequirePackage{lineno}
\documentclass[twocolumn,a4paper,amsmath,amssymb]{revtex4}
\usepackage[urlcolor=blue, hyperindex, colorlinks, bookmarks=true]{hyperref}
\usepackage{lipsum}
\usepackage{times}
\usepackage{amsmath}
\usepackage{braket}
\usepackage{xcolor}
\usepackage{dsfont}
\usepackage{graphicx}
\usepackage{amssymb}
\usepackage{amsthm}
\usepackage{epstopdf}
\usepackage{mathrsfs}
\usepackage{enumerate}
\usepackage{xypic}

\renewcommand{\eqref}[1]{Eq.~(\ref{#1})}

\newcommand{\figref}[1]{Fig.~\ref{#1}}

\newcommand{\removedD}[1]{{\color{gray}{#1}}}
\renewcommand{\removedD}[1]{{}} 

\newcommand{\CEE}[1]{{\color{black}{#1}}}
\newcommand{\CEEE}[1]{{\color{black}{#1}}}
\newcommand{\CEEEE}[1]{{\color{black}{#1}}}

\renewcommand{\eqref}[1]{Eq.~(\ref{#1})}

\newcommand{\appref}[1]{\hyperref[#1]{Appendix~\ref*{#1}}}
\newcommand{\tabref}[1]{\hyperref[#1]{Table~\ref*{#1}}}

\DeclareMathOperator{\tr}{tr}

\theoremstyle{plain}

\begin{document}
\title{Exploring Interacting Quantum Many-Body Systems \\by Experimentally Creating Continuous Matrix Product States in Superconducting Circuits}
\author{C. Eichler$^{1}$\footnote{Current address: Department of Physics, Princeton University, Princeton NJ 08544, USA. Correspondence to ceichler@princeton.edu}, J. Mlynek$^1$, J. Butscher$^1$, P. Kurpiers$^1$,
K. Hammerer$^{2,3,4}$, T.~J.~Osborne$^2$, A. Wallraff$^1$}
\affiliation{$^1$ Department of Physics, ETH Z\"urich, CH-8093, Z\"urich, Switzerland}
\affiliation{$^2$ Institute for Theoretical Physics, Leibniz University, 30167 Hannover, Germany}
\affiliation{$^3$ Institut f\"ur Gravitationsphysik, Leibniz Universit\"at, 30167 Hannover, Germany}
\affiliation{$^4$ Max-Planck Institut f\"ur Gravitationsphysik (Albert-Einstein Institut), 30167 Hannover, Germany}
\begin{abstract}
Improving the understanding of strongly correlated quantum many body systems such as gases of interacting atoms or electrons is one of the most important challenges in modern condensed matter physics, materials research and chemistry.
Enormous progress has been made in the past decades in developing both classical and quantum approaches to calculate, simulate and experimentally probe the properties of such systems. In this work we use a combination of classical and quantum methods to experimentally explore the properties of an interacting quantum gas by creating experimental realizations of continuous matrix product states -– a class of states which has proven extremely powerful as a variational ansatz for numerical simulations.
By systematically preparing and probing these states using a circuit quantum electrodynamics (cQED) system we experimentally determine a good approximation to the ground-state wave function of the Lieb-Liniger Hamiltonian, which describes an interacting Bose gas in one dimension. Since the simulated Hamiltonian is encoded in the measurement observable rather than the controlled quantum system, this approach has the potential to apply to exotic models involving multicomponent interacting fields. Our findings \CEE{also} hint at the possibility
of experimentally exploring general properties of matrix product states and entanglement theory.
The scheme presented here is applicable to \CEEEE{a broad range of} systems exploiting strong and tunable light-matter interactions.
\end{abstract}
\maketitle
Progress in revealing relations between solid state physics and quantum information theory has constantly extended the range of quantum many body problems which are tractable with classical computers. One such successful approach is the density matrix renormalization group (DMRG), which was introduced by White in 1992 \cite{White1992} and since then developed into \CEEEE{a} leading method for numerical studies of strongly interacting one dimensional lattice systems \cite{Schollwoeck2005}.
Later it was realized that the DMRG can be interpreted as a variational optimization over matrix product states (MPS) \cite{Ostlund1995,Dukelsky1998}. The class of matrix product states \cite{Verstraete2008,Landau2015} naturally incorporates an area law for the entanglement entropy \cite{Hastings2007} and is thus ideally suited to parameterize many-body states with finite correlation length \cite{Eisert2010}.
\CEEEE{An interesting connection between the MPS formalism and open quantum systems has recently been discovered \cite{Schoen2005a}, which led to the suggestion of using the high-level of experimental control achievable over open cavity QED systems to create continuous matrix product states \cite{Verstraete2010} as itinerant radiation fields for the purpose of quantum simulations \cite{Osborne2010,Barrett2013}.} In this letter we provide experimental evidence that this concept is indeed capable of determining the properties of strongly correlated quantum systems and offers promising perspectives, complementary to existing digital and analog quantum simulation approaches \cite{Buluta2009} explored with trapped atoms and ions \cite{Greiner2002,Friedenauer2008,Kim2010,Lanyon2011,Brantut2012}.
\begin{figure}[bp!]
\includegraphics[scale=0.962]{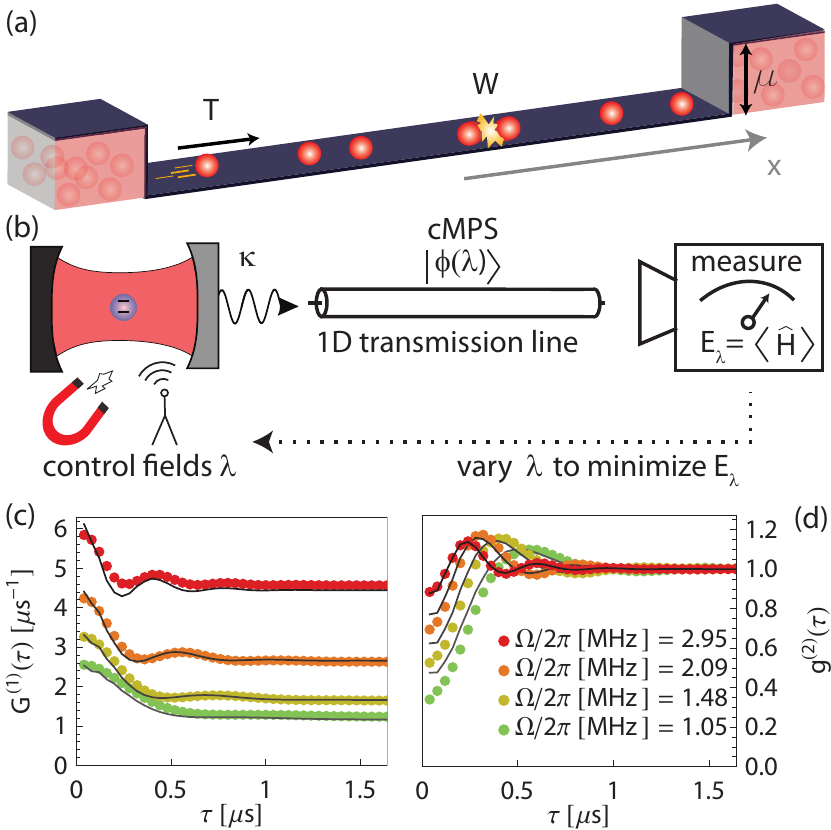}
\caption{
{Schematic of the interacting Bose gas and the principle of the quantum variational algorithm}.
{(a)}, Bosonic particles with kinetic energy $\hat{T}$ are propagating in one dimension along the $x$ axis.
Repulsion between particles mediates an interaction energy $\hat{W}$. The particle density $\rho$ of the gas is controlled by the chemical potential $\mu$.  {(b)},
We experimentally simulate the ground-state of $\hat{H}$ by employing a variational minimization procedure. A tunable cavity QED system is used to generate radiation fields emulating continuous matrix product states $|\phi({\bf \lambda})\rangle$ in a 1D transmission line. The average energy $E_{\lambda} = \langle\phi({\bf \lambda})|\hat{H}|\phi({\bf \lambda})\rangle$ of the simulated Hamiltonian is experimentally determined from measured correlation functions. External control fields ${\bf \lambda}$ are used as variational parameters. {(c),(d)}, Examples of first-order $G^{(1)}(\tau)$ and second-order  $g^{(2)}(\tau)$ correlation functions measured (dots) in superconducting circuits and simulated (solid lines) using a master equation approach for the indicated drive rates $\Omega$, effective anharmonicity $\alpha/2\pi = 5.2\,$MHz, and cavity decay rate $\kappa/2\pi=2.2\,$MHz.
}
\label{fig:1}
\end{figure}
\begin{figure*}
\centering
\includegraphics[scale=1.051]{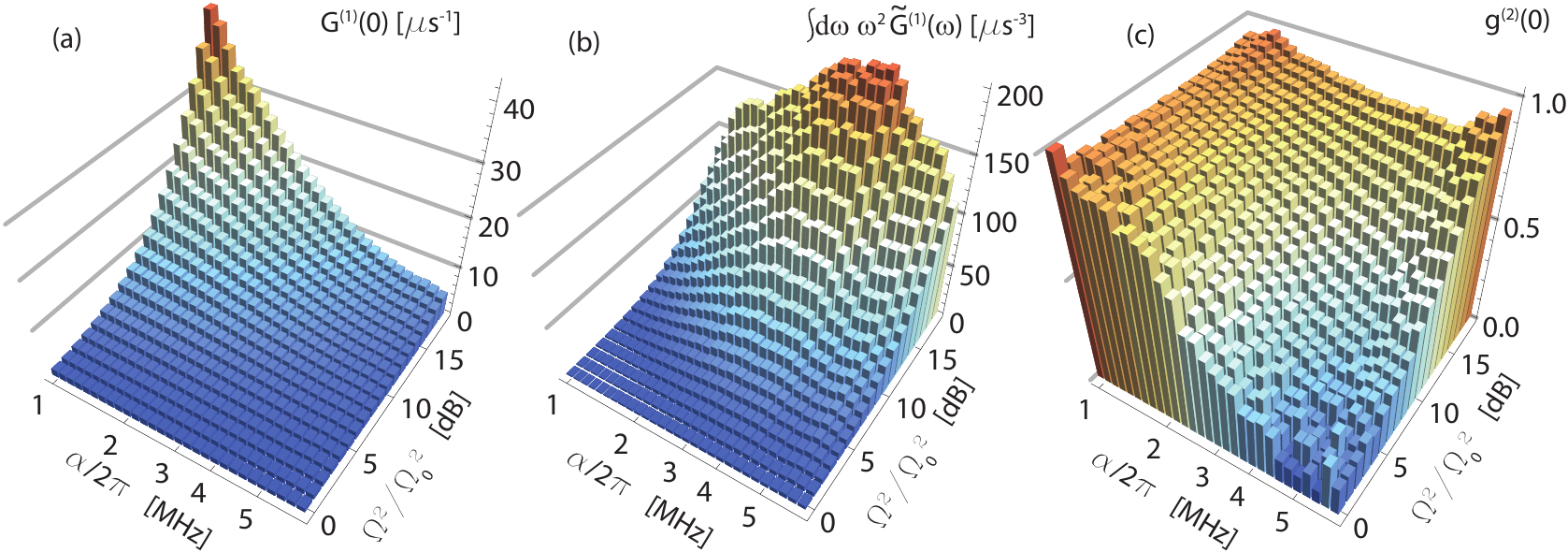}
\caption{
{Measured correlations in the variational space spanned by the anharmonicity $\alpha$ and the drive rate $\Omega$.}
{(a)}  Average photon flux $G^{(1)}(0)$ proportional to the potential energy $\langle\hat{N}\rangle$, {(b)} $\int {\rm d}\omega \omega^2 \tilde{G}^{(1)}(\omega)$ proportional to the kinetic energy $\langle \hat{T}\rangle$, and {(c)} second-order correlator $g^{(2)}(0)$ proportional to the interaction energy $\langle\hat{W}\rangle$.  The smallest drive rate is $\Omega_{0}/2\pi = 0.37\,$MHz.
}
\label{fig:fig2}
\end{figure*}

\CEEEE{In clear distinction to previous experiments, here, we simulate a continuous quantum field theory rather than a lattice model.
In particular, we study the ground-state of the Lieb-Liniger model $\hat{\mathcal{H}} = \int {\rm d}x \hat{H} = \int {\rm d}x (\hat{N}+ \hat{T} + \hat{W})$ \cite{Lieb1963}, which describes
a gas of interacting bosons confined in a one-dimensional continuum \cite{Paredes2004}, as schematically depicted in \figref{fig:1}a.}
Here, $\hat{N} = -\mu \hat{\psi}_x^\dagger \hat{\psi}_x$ is the potential energy, $\hat{T}=\partial_x \hat{\psi}_x^\dagger \partial_x \hat{\psi}_x$ is the kinetic energy of particles,  and $\hat{W}=v(\hat{\psi}_x^\dagger)^2 \hat{\psi}_x^2$ is the interaction energy expressed in second quantization by the field operator $\hat{\psi}_x$.
The Lieb-Liniger model has only one intensive parameter $\tilde{v} = v/\rho$, where $\rho=\langle\hat{\psi}_x^\dagger\hat{\psi}_x\rangle$ is the average particle density and $v$ the interaction strength.
The ground-state energy of this model can be calculated analytically using the Bethe ansatz \cite{Lieb1963}. The calculation of two point correlation functions requires the use of numerical methods such as quantum Monte Carlo or DMRG \cite{Verstraete2010}. The fact that this model is well understood makes it an ideal test case to benchmark the as yet unexplored quantum variational algorithm recently proposed by Barrett et al. \cite{Barrett2013}.

In the experiments presented here, we prepare continuous matrix product states $|\phi(\lambda)\rangle$ \cite{Verstraete2010,Barrett2013} as microwave radiation fields propagating along a one-dimensional transmission line, see  \figref{fig:1}b. The radiation fields are generated by an ancillary quantum system -- in our case a tunable circuit QED system \cite{Wallraff2004}~-- which is coupled with rate $\kappa$ to the transmission line. Notably, any radiation field generated in this way is described by a continuous matrix product state with a bond dimension depending on the number of participating ancillary energy levels \cite{Schoen2005a}. We vary the quantum state $|\phi(\lambda)\rangle$ by tuning a set of two external variational parameters $\lambda = (\alpha,\Omega)$. \CEEEE{Here, $\Omega$ is the drive rate of a coherent field applied resonantly to the upper eigenmode of the coupled system and $\alpha$ is the effective anharmonicity of the driven mode}. Tunability of the effective anharmonicity $\alpha$ is experimentally achieved by employing a qubit of which both the frequency and its coupling to the cavity are adjustable in-situ during the experiment \cite{Srinivasan2011}, see \appref{app:VarPar} for details.

\CEEEE{The variational ground-state of the Lieb-Liniger Hamiltonian $\hat{H}$ is found by measuring the expectation value $E_\lambda = \langle\phi(\lambda)|\hat{H}|\phi(\lambda)\rangle =\langle \hat{N}\rangle + \langle \hat{T} \rangle + \langle \hat{W} \rangle$ and minimizing $E_\lambda$ with respect to different variational states $|\phi(\lambda)\rangle$ created in our experiment}. \CEE{The simulated Hamiltonian is thus solely determined by the measurement observable. Any model of which the corresponding expectation values can be measured, is therefore accessible with this approach.} Given the photonic realization of $|\phi(\lambda)\rangle$, the measurement of $E_\lambda$
translates into the measurement of photon correlation functions. Spatial correlations in the field $\hat{\psi}_x$ are mapped onto time correlations in the cavity output field $\hat{a}_{\rm out}(t)$ by identifying $\hat{\psi}_x=\hat{a}_{\rm out}(t=x/s)/\sqrt{s}$, where the scale parameter $s=x/t$ acts as an additional variational parameter \cite{Barrett2013}. Entanglement in the matrix product states thus corresponds to entanglement between photons emitted from the cavity at different times.
According to this correspondence, $E_\lambda$ for the Lieb-Liniger Hamiltonian is given by the first- and second-order correlation functions
$G^{(1)}(\tau)\equiv \langle \hat{a}^\dagger_{\rm out}(\tau) \hat{a}_{\rm out}(0)\rangle$ and $G^{(2)}(\tau)\equiv \langle \hat{a}^\dagger_{\rm out}(0) \hat{a}^\dagger_{\rm out}(\tau) \hat{a}_{\rm out}(\tau)\hat{a}_{\rm out}(0)\rangle
$
\cite{Gardiner1991}.
More specifically, the average kinetic energy $\langle \hat{T} \rangle = s^{-3} \int {\rm d}\omega \omega^2 \tilde{G}^{(1)}(\omega)$ is calculated from the Fourier transform of the first-order correlation function $\tilde{G}^{(1)}(\omega)$, \CEE{the interaction energy is $\langle \hat{W} \rangle = s^{-2}v G^{(2)}(0)$ and the potential energy is given by the average photon flux $\langle \hat{N} \rangle =- s^{-1}\mu G^{(1)}(0)$.}

The presented variational approach thus crucially  relies on the ability to generate and probe a wide range of different (quantum) radiation fields with high efficiency.  For fast and reliable correlation measurements \cite{Lang2011} we have developed a quantum-limited amplifier which allows for phase-preserving amplification at large bandwidth and high dynamic range \cite{Eichler2014a}.
The examples of measured correlation functions shown in \figref{fig:1}c-d illustrate their dependence on the drive rate $\Omega$ at constant $\alpha$. While $G^{(1)}(\tau=0)$ equals the total average photon flux $\langle a_{\rm out}^\dagger a_{\rm out}\rangle$, the limit $G^{(1)}(\tau\rightarrow\infty)$ is proportional to the square of the coherence of the field $|\langle a_{\rm out} \rangle|^2$. Therefore, $G^{(1)}$ increases with drive rate due to the enhanced photon production rate.  \CEE{The normalized second-order correlation functions $g^{(2)}(\tau) \equiv G^{(2)}(\tau)/(G^{(1)}(0))^2$} show anti-bunched behavior $(g^{(2)}(0)<1)$ for weak drive and Rabi type oscillations when the drive rate $\Omega$ becomes larger than the decay rate \cite{Lang2011}. Both the measured first-order and second-order correlation functions are in agreement with the results obtained from master equation simulations (black solid lines).

We have measured $\sim10^3$ such correlation functions over a wide range of variational parameters $\Omega$ and $\alpha$ by acquiring data for one week. \CEEEE{Without the employed parametric amplifier the time for measuring this set of data would have been on the order of years because of the exponential scaling between statistical error and correlation order \cite{daSilva2010}}. Based on this collection of time-resolved correlation functions we have evaluated the three relevant terms $G^{(1)}(0)$, $\int {\rm d}\omega \omega^2 \tilde{G}^{(1)}(\omega)$ and $g^{(2)}(0)$ that enter the calculation of $E_\lambda$, see \figref{fig:fig2}. \CEEE{As expected, the average photon flux $G^{(1)}(0)$ (\figref{fig:fig2}a) increases with drive rate $\Omega$ and \CEEEE{is suppressed for increasing anharmonicity $\alpha$}}. The kinetic energy term $ \int {\rm d}\omega \omega^2 \tilde{G}^{(1)}(\omega)$ in panel b is determined by the power spectral density $\tilde{G}^{(1)}(\omega)$. Only the spectral weight of photons generated at finite detuning from the drive frequency ($|\omega|>0$) contributes to the integral. \CEEE{In the Bose gas picture such photons correspond to particles in finite momentum states and therefore carrying kinetic energy}. The rate of scattering events from drive photons into photons with finite $\omega$ increases with drive strength and has a non-trivial dependence on $\alpha$. Finally, the second-order correlator $g^{(2)}(0)$ in \figref{fig:fig2}c clearly reveals the crossover from antibunched radiation ($g^{(2)}(0)\rightarrow 0$) for large $\alpha$ and small $\Omega$ to coherent radiation ($g^{(2)}(0)\rightarrow 1$) when lowering the anharmonicity.
\begin{figure}[b]
\centering
\includegraphics[scale=1]{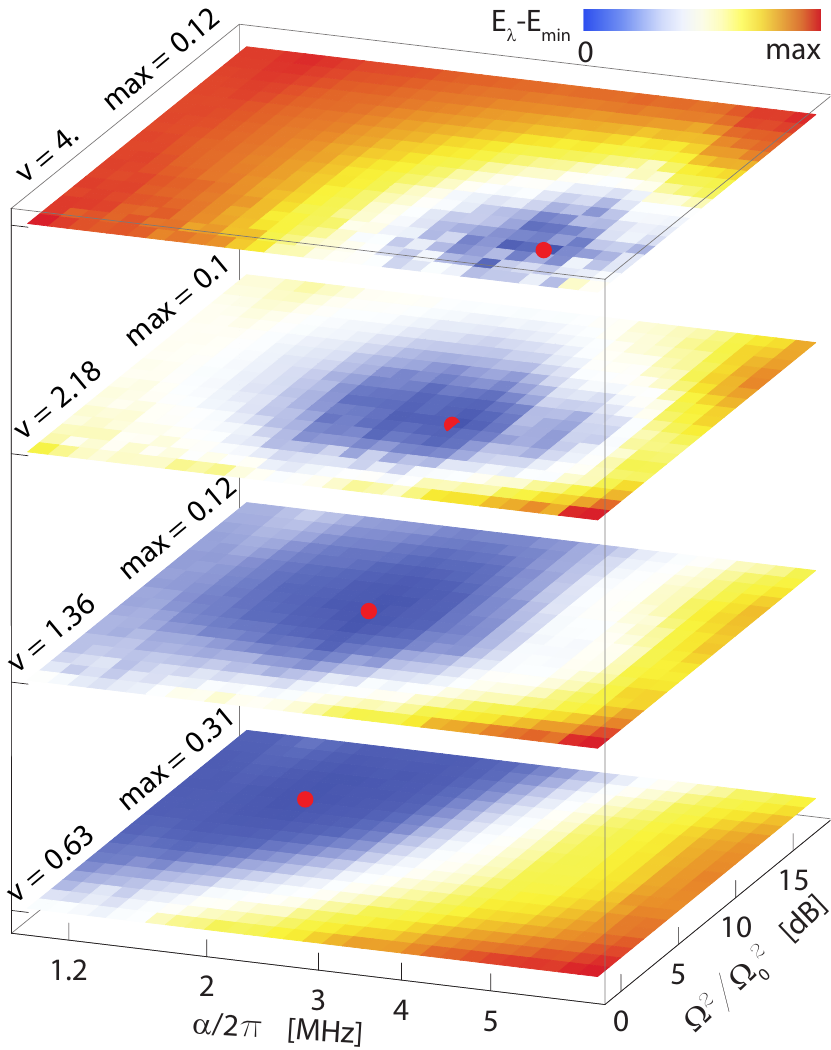}
\caption{{Measured energy landscape for the Lieb-Liniger model.} $E_\lambda$ calculated from the measurement data shown in \figref{fig:fig2} relative to its minimum $E_{\rm min}$ as a function of $\alpha$ and $\Omega$. Interaction strength $v$ increases from bottom to top as indicated.
The color scale is adjusted in each panel such that the maximal value $max$ appears red. The minima from bottom to top are located at $(\Omega/\Omega_0)^2 \in \{{13,10,7,3}\}\,$dB and $\alpha/2\pi \in \{{1.56, 2.25, 3.43, 5.04}\}\,$MHz as indicated by red dots.
}
\label{fig:fig3}
\end{figure}

\CEEEE{Based on the three measured quantities shown in \figref{fig:fig2} and a chosen interaction parameter $v$ we evaluate the energy $E_\lambda$ for all prepared states $|\phi(\lambda)\rangle$. We identify a local minimum in variational space (blue region), which corresponds to the variational ground-state of the Lieb-Liniger model, see \figref{fig:fig3}. When changing the interaction parameter $v$ we find the energy $E_\lambda$ to be minimized by a different set of parameters $(\alpha,\Omega)$ in variational space.
While for large values of $v$ the minimum appears in the anti-bunched region (lower \CEE{right} corner in top panel), the minimum moves to the region where the radiation is mostly classically coherent when weakening the interaction strength (upper \CEE{left} corner in bottom panel). {The maximum and minimum value of $v$ which can be explored in this way is practically limited by the range of variational parameters $\alpha$ and $\Omega$ for which correlation functions have been acquired experimentally.}}

\CEEE{After identifying the variational ground-state for each interaction strength $v$ as the respective minimum in the energy landscape, we further investigate  its properties.} We compare the experimental results with the numerically exact results obtained from a variational matrix product state algorithm executed on a classical computer with bond dimension $D=14$ \cite{Verstraete2010}. We followed the usual convention and rescaled all quantities so that they correspond to a particle density of $\rho=1$ \cite{Verstraete2010}. The Lieb-Liniger ground-state energy density $E_{LL}$, which is the sum of kinetic energy and interaction energy, increases with interaction strength and ideally converges towards the Tonks-Giradeau limit $\underset{\tilde{v}\rightarrow\infty}{\lim}E_{LL}=\pi^2/3$ indicated as a dashed line in  \figref{fig:Supp4}a. Given the small number of variational parameters the experimental data (blue dots) reproduces the characteristic dependence of $E_{LL}$ on $\tilde{v}$ of the exact solution (red solid line) quite well.
\begin{figure*}[t]
\includegraphics[scale=1]{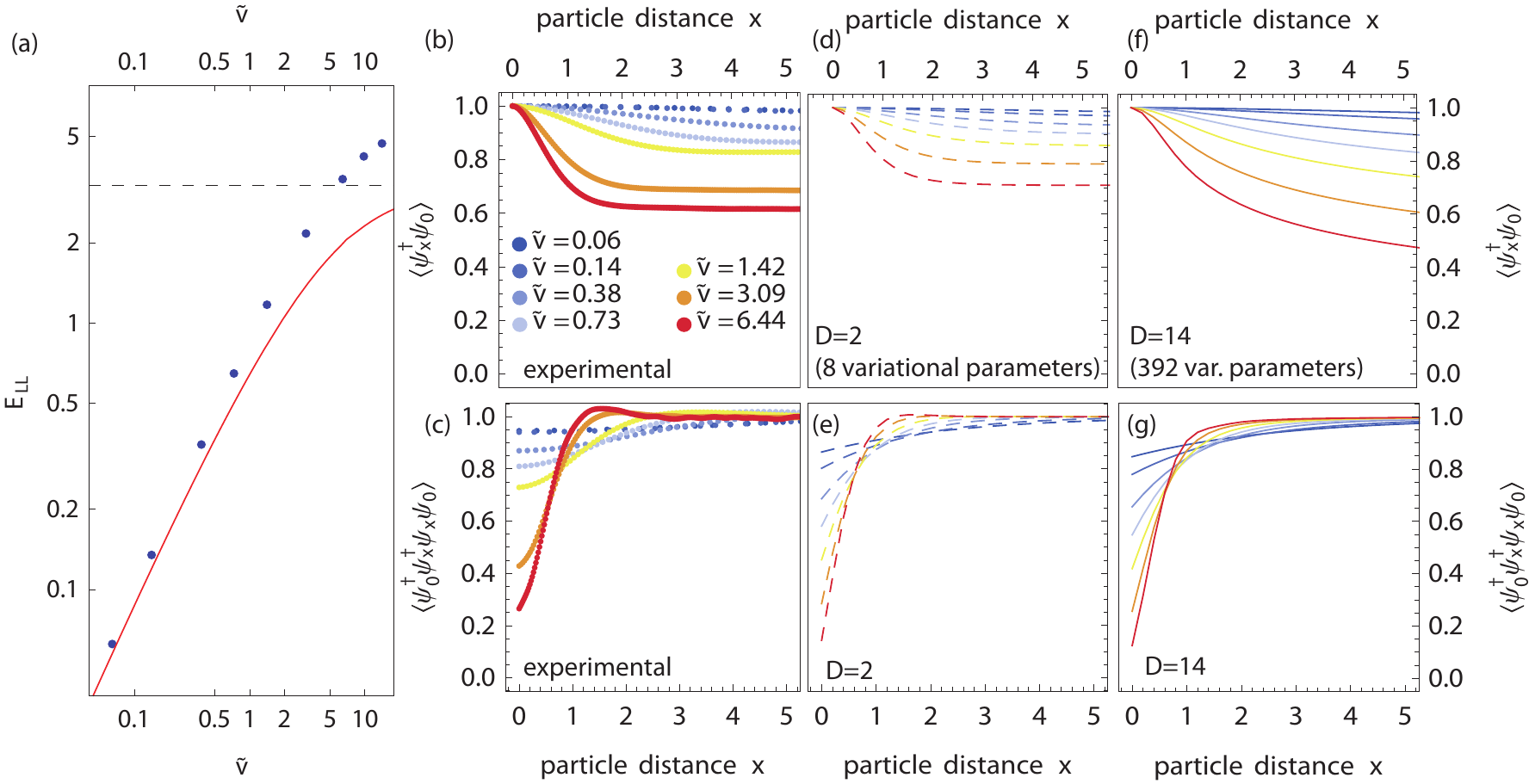}
\caption{{Comparison between experimental simulation and numerical result.} {(a)}  Lieb-Liniger ground-state energy density $E_{LL}$ vs.  interaction parameter $\tilde{v}$ at constant particle density $\rho=1$ on a log/log scale. {(b)-(c)} Experimentally obtained first-order $\langle \hat{\psi}^\dagger_x \hat{\psi}_0\rangle$ and particle-particle correlation functions $\langle \hat{\psi}^\dagger_0 \hat{\psi}^\dagger_x \hat{\psi}_x \hat{\psi}_0\rangle$ for the seven indicated interaction strengths $\tilde{v}$. {(d)-(g)} Corresponding numerical solutions using continuous matrix product states with bond dimensions $D=2$  and $D = 14$.
}
\label{fig:Supp4}
\end{figure*}

Importantly, having physical access to the ground-state \CEEEE{wave functions $|\phi(\lambda)\rangle$} we can also probe quantities beyond the ground-state energy, such as two-point correlation functions.
\CEEEE{First-order correlation functions $\langle \hat{\psi}^\dagger_x \hat{\psi}_0 \rangle$ are obtained from $G^{(1)}(\tau)$ by converting time into spatial coordinates (\figref{fig:Supp4}b).} \CEEE{As expected, we observe a decrease in correlation length with increasing interaction strength $\tilde{v}$.
Due to the absence of spontaneous symmetry breaking in one dimension \cite{Hohenberg1967}, the exact ground state of the Lieb-Liniger model does not exhibit Bose-Einstein condensation. The observed finite limit $\underset{{x}\rightarrow\infty}{\lim}\langle \hat{\psi}^\dagger_x \hat{\psi}_0 \rangle$ is a characteristic feature of  
matrix product states which do not support the $U(1)$ symmetry of the model for finite bond dimensions.}

The nontrivial nature of the ground-state in the presence of interactions also becomes manifest in the particle-particle correlator $\langle \hat{\psi}^\dagger_0 \hat{\psi}^\dagger_x \hat{\psi}_x \hat{\psi}_0\rangle$ shown in panel c. With increasing $\tilde{v}$ particles are more likely to repel each other leading to anti-bunching. Our experiments clearly resolve this \CEEEE{crossover} from a weakly into a strongly interacting Bose gas \CEEEE{by accessing variational wave functions} for interaction parameters $\tilde{v}$ over two orders of magnitude.
While this general behavior is qualitatively well reproduced, an accurate quantitative agreement with the numerical results would require a larger number of \CEE{independent} variational parameters in the experimental realization. This becomes apparent when comparing the experimental results with numerical calculations based on continuous matrix product states of different bond dimensions, $D=2$ and $D=14$, where the number of variational parameters is $2D^2$. Correlation functions simulated with low bond dimension $D=2$ deviate from the exact results ($D=14$) similarly as the measured ones (\figref{fig:Supp4}d-g).

In summary, we have experimentally revealed connections between open quantum systems, the matrix product state variational class, and quantum field theories that can be used for practical quantum simulations. The presented quantum variational algorithm is general in the sense that it can be applied to any one-dimensional quantum field theory. Exploring interacting vector field models seems particularly appealing, since they are difficult to simulate on classical computers. Experimentally, this could be achieved by coupling tunable quantum systems to multiple transmission lines each representing one of the vector field components \cite{Quijandria2014}. Higher accuracy in the simulation will require more variational parameters and quantum systems with more degrees of freedom, which is achievable with tunable superconducting circuits. In this context, the collective dissipation of multiple emitters coupled to the same transmission line at finite distance \cite{vanLoo2013} may turn out very useful. \CEEEE{Extensions of the presented approach may be envisaged to also explore dynamical phenomena and discrete lattice models using experimentally created matrix product states.}

We thank Christoph Bruder, Ignacio Cirac, Tilman Esslinger and Atac Imamoglu for discussions and comments. This work was supported by the European Research Council (ERC) through a Starting Grant, by the NCCR QSIT and by ETHZ. CE acknowledges support by Princeton University through a Dicke fellowship. TJO was supported by the ERC grants QFTCMPS and SIQS, and by the cluster of excellence EXC201 Quantum Engineering and Space-Time Research.
\appendix
\section{Experimental details}
\subsection{Measurement setup, sample fabrication and characterization}
\begin{figure*}[t]
\includegraphics[scale=1]{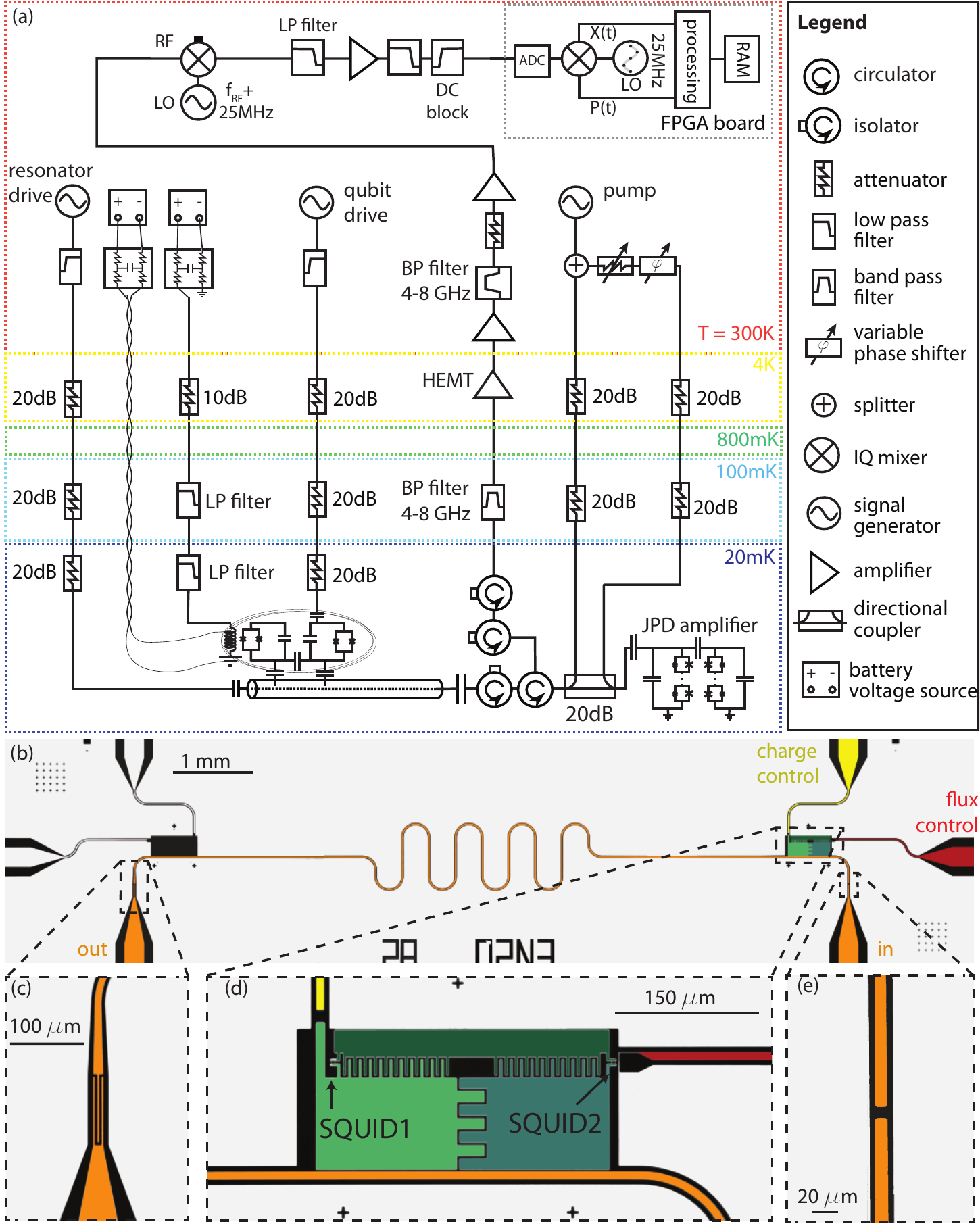}
\caption{{(a)} Schematic of the measurement setup. For details see text. {(b)} Optical micrograph of the sample. The second qubit gap in the left part of the chip is left empty. Enlarged images of the output capacitor {(c)}, of the qubit {(d)}, and of the input capacitor {(e)}  are shown.
}
\label{fig:Supp1}
\end{figure*}
The experiments presented in the main text are performed with a device consisting of a superconducting circuit, see \figref{fig:Supp1}a for details about the experimental setup. The sample (\figref{fig:Supp1}{b}) consists of a $\lambda/2$ transmission line cavity with a resonance frequency $\omega_{\rm res}/2\pi \approx 7.3425\,$GHz of the fundamental mode. The resonator has one output port which dominates the total decay rate $\kappa/2\pi \approx 2.2\,$MHz and one weakly coupled input port $\kappa_{\rm in}/\kappa \approx 0.01$ which is used for coherent driving of the cavity field. The resonator is fabricated using photolithography and reactive ion etching of a Niobium thin film sputtered on a sapphire wafer.  We have fabricated a superconducting qubit (\figref{fig:Supp1}{c}) close to one end of the resonator. Both its transition frequency $\omega_{ge}$ and coupling strength $g$ to the resonator are tunable by varying the magnetic fluxes threading the two SQUID loops \cite{Srinivasan2011,Gambetta2011}. Flux control is achieved by a combination of a superconducting coil mounted on the backside of the sample holder and a local flux line which couples predominantly to one of the two SQUIDs. The currents feeding the coil and the flux line are generated by voltage biased resistors at room temperature. The qubit is fabricated using double-angle evaporation of aluminum on a mask defined by electron beam lithography. The qubit decay and dephasing times are measured to be $T_1\approx1.8\,\mu$s and $T_2\approx1.2\,\mu$s. The anharmonicity of the qubit is $(\omega_{ef}-\omega_{ge})/2\pi \approx -80\,$MHz, where $\omega_{ef}$ is the transition frequency from the first excited state $|e\rangle$ to the second excited state $|f\rangle$ of the qubit.

The sample is mounted on the base plate of a dilution refrigerator cooled down to a temperature of about 20 mK. The qubit and the resonator are coherently driven through attenuated charge control lines. The microwave radiation emitted from the cavity is guided through two circulators to a Josephson parametric dimer (JPD), which provides quantum-limited amplification at large bandwidth and dynamic range \cite{Eichler2014a,Castellanos2008,Vijay2009}. A directional coupler is used to apply and interferometrically cancel the reflected pump field. The amplified signal reflects back from the JPD, passes a bandpass (BP) filter, is further amplified by a high electron mobility transistor (HEMT) amplifier, and is down-converted to an intermediate frequency (IF) of 25 MHz. After low-pass (LP) filtering and IF amplification the down-converted signal is digitized using analog-to-digital conversion (ADC) and further processed with field programmable gate array (FPGA) electronics.
\begin{figure*}[t]
\includegraphics[scale=1]{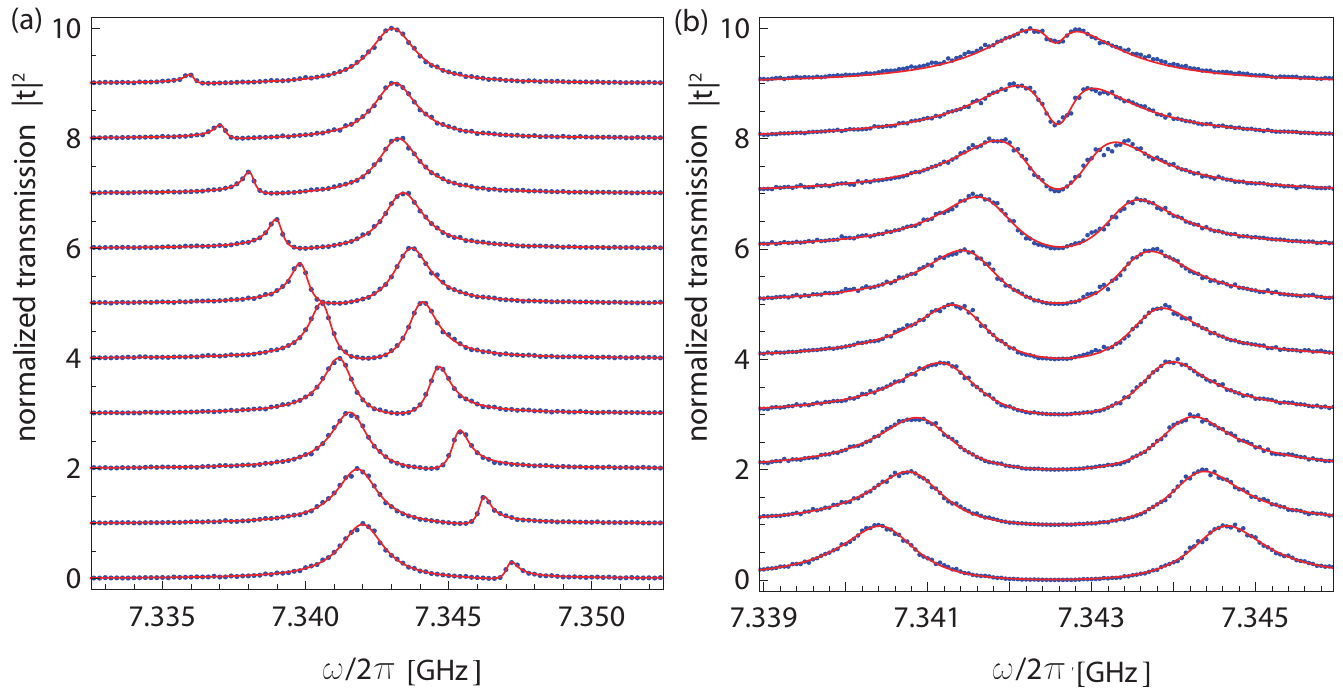}
\caption{{(a)} Measurements and fits of the absolute square of the transmission coefficient $|t|^2$ for varying qubit frequency at approximately constant coupling strength $g/2\pi = 1.75\,$MHz. Individual data traces are offset from each other by one. {(b)} Transmission coefficient $|t|^2$ for varying coupling strength $g$ at constant qubit frequency $\omega_{ge} \approx \omega_{\rm res} $.
}
\label{fig:Supp2}
\end{figure*}
\subsection{Controlling the qubit frequency and the coupling strength}
We characterize the coupled cavity-qubit system by probing the transmission coefficient of the cavity and fitting the data to the absolute square of the expression
\begin{equation} t = \frac{A \kappa}{i(\omega_{\rm res}-\omega) + \frac{g^2}{i(\omega_{ge}-\omega)+\gamma/2}+\kappa/2},
\label{eq:transmission}
\end{equation}
which we obtain from input-output theory for the Jaynes-Cummings model \cite{Gardiner1999}. Here, $\omega$ is the probe frequency, $\gamma$ is the qubit decoherence rate and $A$ is a scaling factor. The probe power is chosen such that the average number of excitations of the coupled resonator qubit system is much smaller than one. In this case the qubit may be approximated by an harmonic oscillator. We determine the qubit detuning $\Delta=\omega_{\rm res}-\omega_{ge}$ and its coupling strength $g$ to the cavity by fitting spectroscopically obtained transmission data to the above model, see \figref{fig:Supp2}. The magnetic fluxes through the qubit SQUID loops and with that the qubit parameters $g$ and $\Delta$ are controlled by a pair of voltages $V_1$ and $V_2$ applied to coil bias resistors.
For small $g$ and $\Delta$ we approximate the relation between $(g,\Delta)$ and $(V_1,V_2)$ by linear equations of the form
\begin{equation}
  \begin{pmatrix} g \\ \Delta \end{pmatrix}
=  \begin{pmatrix} m_{11} & m_{12}  \\ m_{21} & m_{22} \end{pmatrix}
  \begin{pmatrix} V_1-V_{1,0} \\ V_2 - V_{2,0} \end{pmatrix}.
\label{eq:matrix}
\end{equation}
We determine the coupling matrix elements $m_{ij}$ and the offset voltages $V_{1,0}$ and $V_{2,0}$ by recording transmission data for pairs $(V_1,V_2)_k$. For each set of data we extract the corresponding parameters $(g,\Delta)_{k}$ and perform a least-square fit to determine the model parameters $m_{ij}$, $V_{1,0}$ and $V_{2,0}$. By inverting \eqref{eq:matrix} we calculate the voltages $V_1$ and $V_2$ for a given set of desired qubit parameters $(g,\Delta)$. In order to further fine-tune the parameters, we have developed an automated calibration algorithm which minimizes the deviations from desired target values by iteratively measuring, fitting and readjusting the control parameters $V_1$ and $V_2$. With this control procedure we are able to independently set the qubit frequency and the interaction strength. We also use this procedure to monitor and correct for slow qubit frequency drifts occurring during long runs of the experiment.

We demonstrate individual control of qubit parameters by either tuning the qubit frequency for approximately constant coupling strength $g$ (\figref{fig:Supp2}a) or by varying $g$ for  fixed qubit frequency (\figref{fig:Supp2}b).
For all sets of data we have turned on the JPD amplifier and have divided out its frequency dependent gain. For the measurements in \figref{fig:Supp2}b we have kept the qubit resonant with the cavity ($\Delta=0$) and have varied $g$. These measurements demonstrate the ability to tune the system from the fast cavity $(\kappa \gg g\gg\gamma)$ into the strong coupling regime $(g\gg\gamma,\kappa)$ \cite{Srinivasan2011}.
\begin{figure*}
\includegraphics[scale=1]{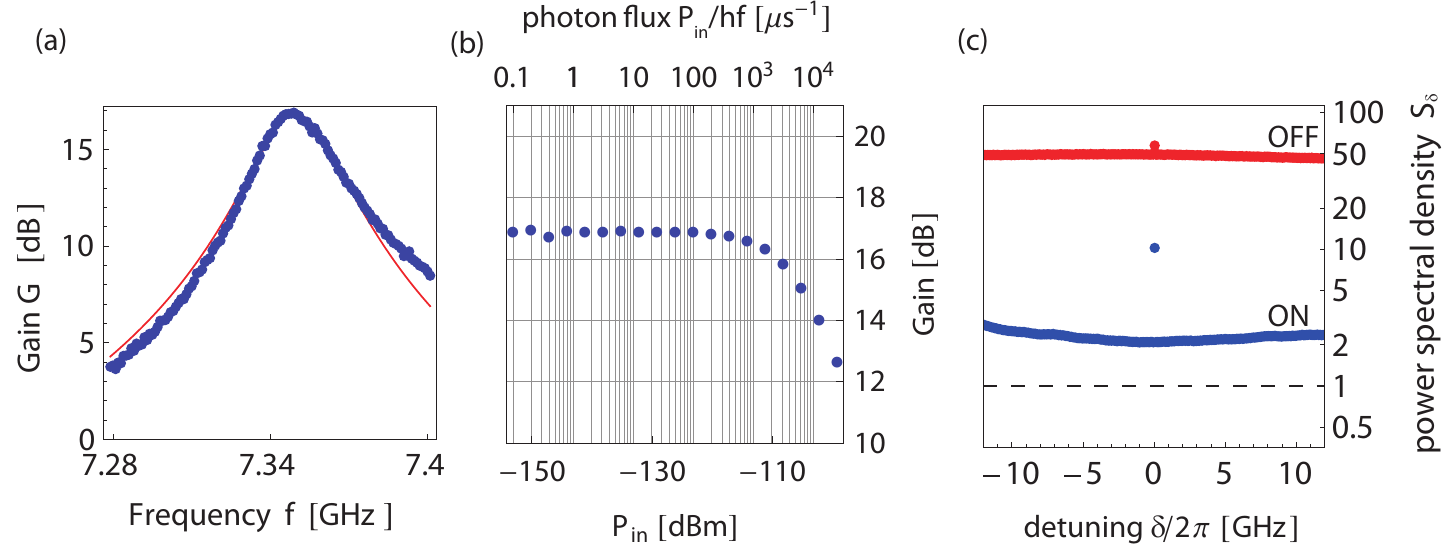}
\caption{{(a)} Measured gain of the JPD amplifier (blue dots) and a Lorentzian fit (solid red line). {(b)} Measured JPD gain as a function of signal power and equivalent photon flux. {(c)} Measured noise power spectral density $S_\delta$ referenced back to the input of the JPD amplifier in units of photons per Hz per second when the JPD is turned on (blue) and when it is turned off (red). The detuning $\delta$ is relative to the frequency of a coherent test tone applied to the JPD input at frequency 7.35 GHz. The dashed line indicates the quantum noise limit for phase-preserving amplification and detection.
}
\label{fig:Supp3}
\end{figure*}
\subsection{Josephson parametric dimer amplifier}
To measure higher order correlation functions efficiently while retaining a high level of linearity we employ a Josephson parametric dimer (JPD) amplifier.
For details about the operational principles of the JPD we refer the reader to \cite{Eichler2014a}. The gain $G$ measured vs. signal frequency $f$ is approximately described by a Lorentzian function (\figref{fig:Supp3}). The amplifier bandwidth at full width half maximum is approximately 35 MHz. In order to increase dynamic range, we have chosen a moderate maximum gain of 17 dB. A measurement of the gain as a function of signal power $P_{\rm in}$ results in a 1 dB compression point at about -107 dBm which corresponds to a photon flux of 3000 $\mu {\rm s}^{-1}$, see \figref{fig:Supp3}b. The largest photon flux which is generated in the described experiments is less than 50 $\mu {\rm s}^{-1}$. The JPD amplifier is thus far away from its compression point for all measured correlation functions. The improvement in detection efficiency becomes apparent when measuring the noise power spectral density $S_{\delta}$ in units of photons per Hz per second when the JPD amplifier is turned ON and when it is turned OFF.  The effective noise level, which is referenced back to the input of the JPD amplifier is decreased by more than an order of magnitude when it is turned on. The scaling of $S_{\delta}$ is based on a comparison between the frequency dependent gain $G_{\delta}$ and the JPD amplifier noise \cite{Eichler2011a}. The deviation from the quantum limit is due to the additional noise from the following HEMT amplifier which is comparable to the noise at the output of the JPD. The equivalent detection efficiency of the amplification chain is $\eta_{\rm amp} = 1/S_\delta \approx 50\%$.
\subsection{Calibration of drive rate and output power}
We have calibrated the total gain of the detection chain including all cable losses in order to reference the measured photon flux back to the output of the cavity.
\begin{figure}[b]
\includegraphics[scale=1]{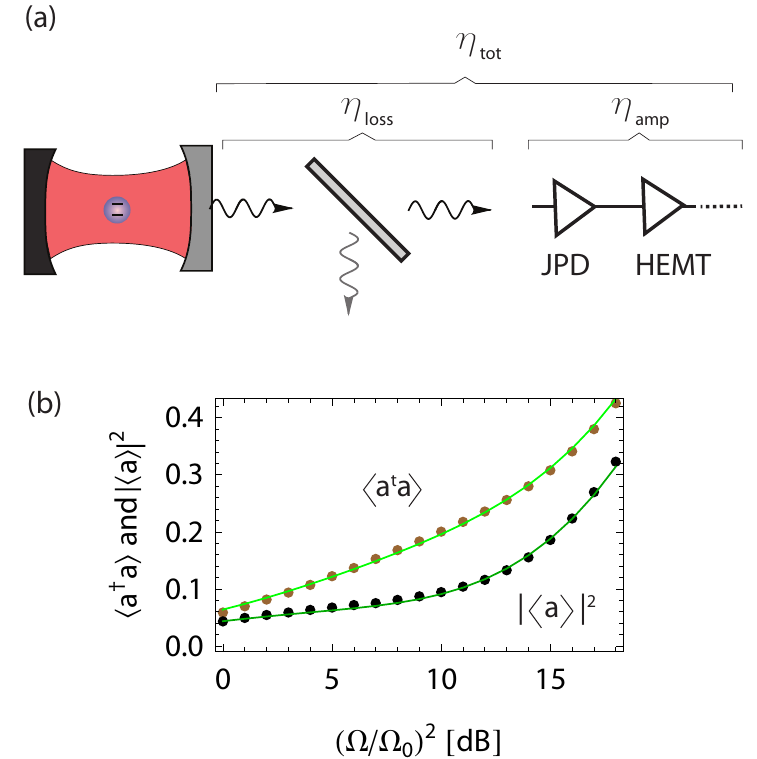}
\caption{{(a)} The detection efficiency of the cavity output field is typically limited by radiation loss, schematically represented as a beamsplitter with finite transmittivity $\eta_{\rm loss}$, and by noise $S_\delta = \eta_{\rm amp}^{-1}$ added in the amplification chain. The total detection efficiency is given by the product $\eta_{\rm tot}=\eta_{\rm loss} \eta_{\rm amp}$. {(b)} Measurement (points) and fit (solid line) of the cavity photon number and absolute square of the coherent amplitude for the coupled cavity-qubit system driven through the cavity input port at rate $\Omega$. The smallest drive rate is $\Omega_{0}/2\pi = 0.37\,$MHz.
}
\label{fig:Supp4}
\end{figure}
\begin{figure*}[t]
\includegraphics[scale=1]{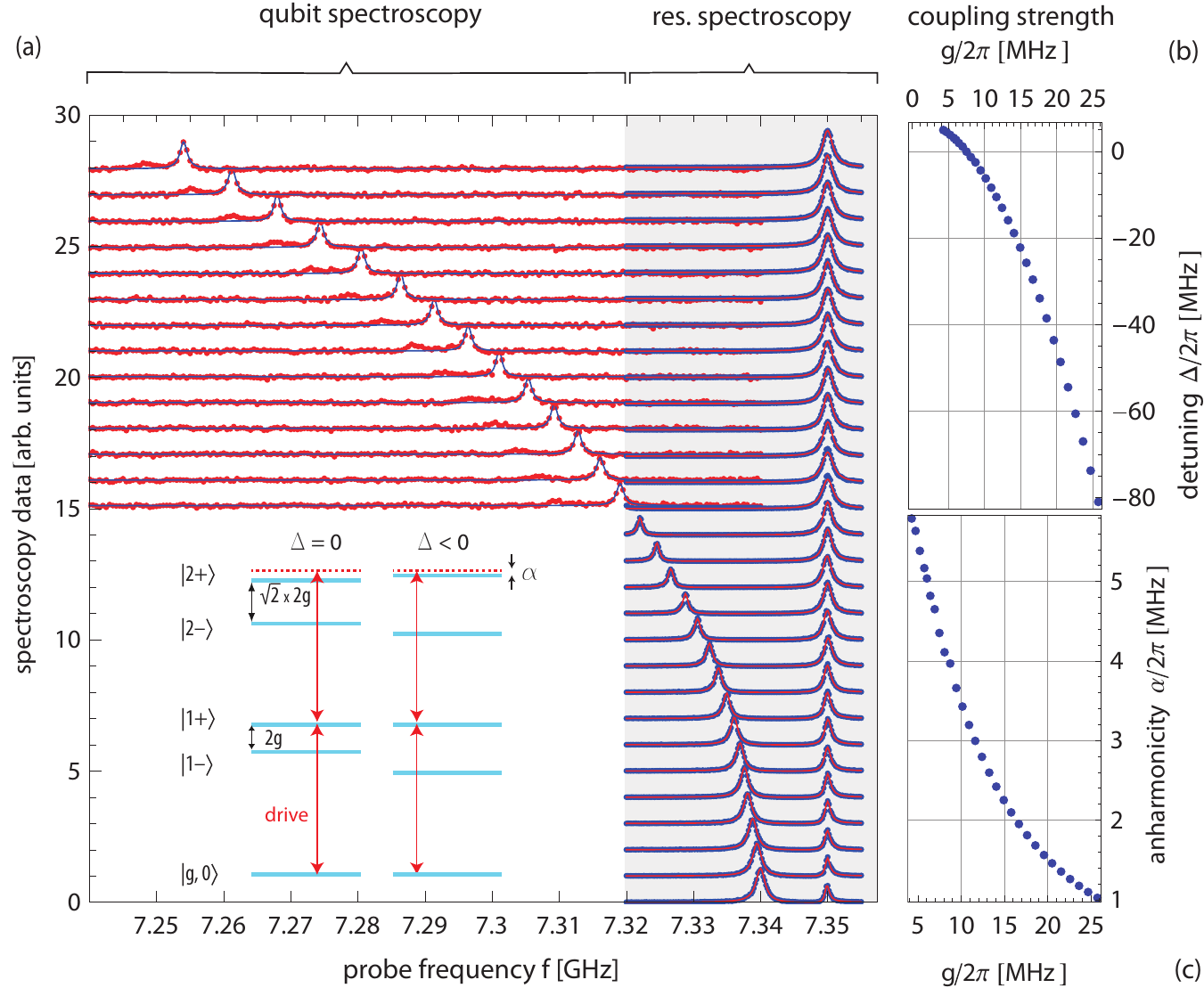}
\caption{{(a)} Spectroscopy data for all values of $g$ used in the quantum simulation experiment. The blue data points on gray background are obtained using resonator transmission measurements. The red data points are obtained using qubit spectroscopy measurements. The relative scaling and offsets are adjusted to display the data in the same plot. The inset shows the energy levels $|n\pm\rangle$ for the Jaynes Cummings model. { (b)} Pairs of $(g,\Delta)$ extracted from the data shown in ${a}$ using the fitting routine described before. {(c)} Effective anharmonicity of the upper branch of the Jaynes Cummings ladder calculated from the pairs $(g,\Delta)$ shown in {b}.}
\label{fig:Supp5}
\end{figure*}
Calibrating the total gain of the detection chain is equivalent to calibrating its detection efficiency $\eta_{\rm tot}$. The detection efficiency is typically limited by losses between the cavity and the first amplifier ($\eta_{\rm loss}$) and by noise added during the amplification process ($\eta_{\rm amp}$), see \figref{fig:Supp4}.  We compare the nonlinear response of the coupled cavity-qubit system with master equation simulations to perform this calibration. We bias the qubit with parameters $(\Delta,g)/2\pi = (3,5.7)\,$MHz and apply a drive field  to the input port of the cavity at rate $\Omega$ and resonant with the frequency $\omega_{+} = \omega_{ge} + \Delta/2 + \sqrt{g^2 + \Delta^2/4} = 2\pi \times 7.35\,$GHz of the upper Jaynes-Cummings doublet state. For these settings we measure the coherent photon flux $\kappa |\langle a \rangle|^2$ and the total photon flux $\kappa |\langle a^\dagger a \rangle|^2$ emitted from the cavity for different drive rates $\Omega$. We fit these data sets to the results obtained from master equation simulations leaving the the total gain factor of the measurement chain and the absolute drive rate incident to the sample as free parameters (\figref{fig:Supp4}b).
The equivalent detection efficiency resulting from this fit is equal to the inverse of the scaled noise level and found to be $\eta_{\rm tot} = \eta_{\rm loss}\eta_{\rm amp} = 0.27$. %
Together with the estimate for the efficiency of the amplification chain $\eta_{\rm amp}$  stated in the previous section, we extract a radiation loss of $1-\eta_{\rm loss} = 0.46$ between the cavity and the JPD, which is reasonable given the components and cables connecting the two stages.
\subsection{Drive scheme and variational parameters}
\label{app:VarPar}
In the original proposal for simulating the Lieb-Liniger model with cavity QED it has been suggested to keep the qubit resonant with the cavity and use the qubit drive power $\Omega_q$ and the coupling strength $g$ as two variational parameters. We have experimentally realized this scheme and found that in the limit of small coupling strengths $g$ the total emission rate becomes extremely small which in turn limits the signal to noise ratio. This is because the effective emission bandwidth scales like $g^2/\kappa$  when $g<\kappa$ and thus decreases quadratically with $g$.  We have therefore developed an alternative scheme for which the photon emission rate remains proportional to $\kappa$ even in the limit of small $g$. We have therefore made use of the ability to tune both the qubit frequency and the coupling strength. Rather than keeping the qubit at fixed frequency we adjust for each value of $g$ the detuning $\Delta$ such that $\omega_+$ remains at constant frequency resonant with the drive frequency $\omega_+ = \omega_d = 2\pi \times 7.35\,$GHz. The spectroscopy data for the qubit bias points used in the quantum simulation experiment are shown in \figref{fig:Supp5}a and demonstrate constant $\omega_+$ over the entire range of  coupling strengths. In order to keep $\omega_+$ constant, we compensate the larger splitting when increasing $g$  by tuning the qubit further away from the cavity, see \figref{fig:Supp5}b.

For this specific tuning scheme we find that the effective anharmonicity $\alpha$ of the upper Polariton ladder decreases with increasing coupling strength $g$, as shown in \figref{fig:Supp5}c. To illustrate this effect we show a schematic drawing of the energy levels $|n\pm\rangle$ of the Jaynes Cummings model for the resonant case ($\Delta=0$) and for the case of finite qubit detuning $\Delta<0$. The inverse proportionality between anharmonicity $\alpha$ and coupling strength $g$ illustrates the appearance of anti-bunched behavior for the small values of $g$ and the observed coherent radiation for large $g$ values for this bias scheme, see Fig.~2 of the main text. The drive rate $\Omega$ of a coherent field applied to the cavity input port acts as a second variational parameter for the quantum simulation.
\subsection{Measurement of correlation function and master equation simulation}
We employ fast real-time signal processing performed with an FPGA at a clock rate of 100 MHz for the measurement of time-resolved correlation functions. The cavity output field is processed, as described in section~1A. After digitization we multiply the sampled voltages with digital sine and cosine waves of frequency $\omega_{\rm IF}/2\pi = 25\,$MHz to obtain the quadrature components $I(t)$ and $Q(t)$, respectively. We then apply an FIR filter with
an effective bandwidth of $\Gamma/2\pi = 10\,$MHz to the quadratures in time-domain to obtain the filtered quadratures $\tilde{I}(t)$ and $\tilde{Q}(t)$, which in the following we write as the single complex valued amplitude $S(t) = \tilde{I}(t) + i \tilde{Q}(t)$. To measure the first-order correlation function in $S(t)$ we take the discrete Fourier transform ($\mathcal{F}$) of $M$ time traces $S_i(t)$ of length 10.24 $\mu$s, multiply with their complex conjugate, and average
$$\Gamma^{(1)}_{}(\tau) = \mathcal{F}^{-1}\left[\frac{1}{M}\sum_{i=1}^{M} \mathcal{F}[S_i(t)]\mathcal{F}^{*}[S_i(t)] \right].$$
Similarly we extract the second-order correlation function by calculating the absolute square of $S(t)$ before Fourier transforming
$$\Gamma^{(2)}_{}(\tau) = \mathcal{F}^{-1}\left[\frac{1}{M}\sum_{i=1}^{M} \mathcal{F}[S^*_i(t)S_i(t)]\mathcal{F}^{*}[S^*_i(t)S_i(t)] \right].$$
We record each of these quantities with the drive field turned on, giving $\Gamma^{(1)}_{\rm ON}(\tau), \Gamma^{(2)}_{\rm ON}(\tau)$, and with the drive turned off, giving $\Gamma^{(1)}_{\rm OFF}(\tau), \Gamma^{(2)}_{\rm OFF}(\tau)$. To avoid effects due to slow drifts we alternate between all four measurements every 12.5 $\mu$s.
As explained in detail in reference~\cite{daSilva2010} and as demonstrated in many experiments since then \cite{Bozyigit2011,Lang2011,Hoi2012b}, we can use these four measurements to extract the correlation functions $G^{(1)}(\tau)=\kappa \langle a^\dagger(\tau)a(0)\rangle$ and $g^{(2)}(\tau)= \langle a^\dagger(0)a^\dagger(\tau) a(\tau)a(0)\rangle/\langle a^\dagger a\rangle^2$ of the output field of the cavity. In these expressions, $a$ ($a^\dagger$) is the annihilation (creation) operator of the intra-cavity field. For the cases in which the average photon number is small $(\langle a^\dagger a \rangle)\ll1$ we find $g^{(2)}(0)$ values which are systematically smaller than the corresponding master equation simulation. We attribute this to a weak thermal background radiation during the off measurements which we correct for \cite{Lang2013}. Good agreement between the measured and simulated correlation functions is found when correcting for a thermal photon flux of $n_{\rm th} \approx 0.03/\mu s$ in the detection band.

We compare these measurements with correlation functions obtained from master equation simulations. For these simulations we describe the system by the Hamiltonian
\begin{eqnarray}
H_{\rm sys}/\hbar &=&  (\omega_{\rm res}-\omega_{d})a^\dagger a + (\omega_{ge}-\omega_{d})b^\dagger b
\nonumber
\\
&& + \frac{\alpha_{q}}{2} (b^\dagger)^2 b^2 + g(a^\dagger b + a b^\dagger)
\end{eqnarray}
expressed in a frame rotating at the drive frequency $\omega_{d}/2\pi = 7.35\,$GHz. Here, $b$ and $b^\dagger$ are annihilation and creation operators for an excitation of the transmon. In addition to that, we account for qubit decay $\gamma$, qubit dephasing $\gamma_{\phi}$ and resonator emission $\kappa$ with standard Lindblad terms. Simulations are run in a Hilbert space including 6 resonator and 3 transmon levels.
In order to account for the finite detection bandwidth when simulating the second-order correlation function we employ the techniques described in \cite{delValle2012}. In this approach we introduce an ancillary mode $c$ of frequency $\omega_d$, which is weakly coupled with rate $\epsilon/2\pi=20\,$kHz to the cavity and decays with a rate equal to the detection bandwidth $\Gamma$. The second-order correlation function in $c$ is then simulated based on the total Liouvillian and taken as an estimate for the filtered correlation function of mode $a$.
\section{Theoretical aspects and data analysis}
\subsection{Calculation of the Lieb-Liniger energy from correlation functions}
\label{sec:min}
The expectation value to be minimized
$\langle \hat{H}\rangle = \langle \hat{T} \rangle + \langle \hat{W}\rangle +\langle \hat{N} \rangle$  is composed of the kinetic energy of the bosons  $\langle \hat{T}\rangle$, its interaction energy  $\langle \hat{W} \rangle$ and the potential energy  $\langle \hat{N} \rangle$. According to the correspondence between the field operator $\hat{\psi}_x$ and the time-dependent radiation field $\hat{a}_{\rm}(t)=\sqrt{s} \hat{\psi}_{x=st} $, each of these expectation values is proportional to a specific measured correlation function
\begin{eqnarray}
\langle \hat{T} \rangle & = & \frac{1}{s^3}\int{\rm d}\omega\tilde{G}^{(1)}(\omega)\omega^2,
\nonumber
\\
\langle \hat{W} \rangle & = & \frac{v}{s^2}G^{(2)}(0),
\nonumber
\\
\langle \hat{N} \rangle & = & -\frac{\mu}{s} G^{(1)}(0) = - \mu \rho.
\label{eq:energies}
\end{eqnarray}
Here, $\tilde{G}^{(1)}(\omega)\equiv \mathcal{F}[\tilde{G}^{(1)}(\tau)]$ is the Fourier transform of the first-order correlation function normalized such that $\int{\rm d}\omega\tilde{G}^{(1)}(\omega)= {G}^{(1)}(0)$. The energy terms in \eqref{eq:energies} thus explicitly depend on the scaling parameter $s$, which is to be treated as an additional variational parameter. We explicitly minimize $\langle H \rangle$ with respect to $s$ by solving $\frac{\partial}{{\partial}s}\langle H \rangle =0$, which results in
\begin{widetext}
\begin{eqnarray}
s = \frac{3 \int{\rm d}\omega\tilde{G}^{(1)}(\omega)\omega^2}{- v G^{(2)}(0) + \sqrt{\left( G^{(2)}(0)\right)^2 v^2 + 3  \mu {G}^{(1)}(0) \int{\rm d}\omega\tilde{G}^{(1)}(\omega)\omega^2}}.
\label{eq:sMin}
\end{eqnarray}
\end{widetext}
Correspondingly we obtain an explicit expression for
$E_{\lambda}=\langle \hat{H} \rangle$, which only depends on the measured correlation functions and the model parameters $\mu,v$.
We find the variational ground state  for a given set of model parameters $(v,\mu)$ by minimizing $E_\lambda$ with respect to $\alpha$ and $\Omega$.
\subsection{Scaling transformation}
We follow the usual convention and study the Lieb-Liniger ground state subject to the constraint that its particle density $\rho$ is equal to one. Using the procedure described in Sec.~\ref{sec:min} we find a ground state which generally does not obey this property. We therefore apply a scale transformation by adjusting the chemical potential $\mu$ such that $\rho\rightarrow1$. Under the following transformation
$$(\mu, v)\rightarrow (\tilde{\mu},\tilde{v})=(\mu y^2, v y),$$
the ground state remains invariant up to a change in the parameter $s \rightarrow \tilde{s} = s/y$, which immediately follows from \eqref{eq:sMin}. Any variational ground state can therefore be transformed into another ground state satisfying $\rho=1$, by choosing $y$ appropriately. We apply the following procedure to perform this scale transformation:
\begin{itemize}
  \item Chose parameter $v$, set $\mu=1$, and find the set of variational parameters $(s_{\rm min},\Omega_{\rm min},\alpha_{\rm min})$ minimizing $E_\lambda$.
  \item Evaluate the particle density $\rho = G^{(1)}(0)/s_{\rm min}$ at this minimum.
  \item Calculate the new chemical potential $\tilde{\mu}=\mu/\rho^2 $ and the new interaction parameter $\tilde{v} = v/\rho$. The new scaling parameter becomes $\tilde{s}_{\rm min}={s}_{\rm min} \rho = G^{(1)}(0) $.
  \item The variational parameters $(\tilde{s}_{\rm min},\Omega_{\rm min},\alpha_{\rm min})$ specify the variational ground state for the model with interaction strength $\tilde{v}$ and unit particle density.
\end{itemize}
Ground states for different interaction parameters are obtained by starting the above procedure with a different value for $v$.
The Lieb-Liniger energy $E_{LL}$ at interaction strength $\tilde{v}$ (Fig.~4a of the main text) is given by
\begin{eqnarray}
E_{LL} &=& \langle \hat{T} \rangle + \langle \hat{W} \rangle
\nonumber
\\
&& \hspace{-10mm} =  \frac{1}{\tilde{s}_{min}^3}\int{\rm d}\omega\tilde{G}^{(1)}(\omega)\omega^2 +  \frac{\tilde{v}}{\tilde{s}_{min}^2}g^{(2)}(0)\left(G^{(1)}(0)\right)^2
\nonumber
\\
&&\hspace{-10mm}  = {\left(G^{(1)}(0)\right)^{-3}}\int{\rm d}\omega\tilde{G}^{(1)}(\omega)\omega^2 +  {\tilde{v}}g^{(2)}(0).
\end{eqnarray}
Correlation functions for the Lieb-Liniger model are directly obtained from the measured correlation functions by identifying
\begin{eqnarray}
\langle \hat{\psi}_x \hat{\psi}_0 \rangle = \frac{G^{(1)}(\tau=x/\tilde{s}_{min})}{G^{(1)}(0)},
\end{eqnarray}
and analogously for the second-order correlation function
\begin{eqnarray}
\langle \psi^\dagger_x \psi^\dagger_0 \psi_0 \psi_x\rangle = g^{(2)}(\tau=x/\tilde{s}_{min}).
\end{eqnarray}
\subsection{Numerically exact solution}
The exact solution \cite{Lieb1963a,Lieb1963}, via the Bethe ansatz, of the Lieb-Liniger model at
unit density is only possible for one specific value of the interaction parameter, namely $v=2$. In order to calculate
properties of the model at unit density for other values of $v$ it is necessary to take recourse to numerical methods.
We exploited a variational method over \emph{translation invariant continuous matrix
product states} (cMPS) \cite{Verstraete2010,Osborne2010,Haegeman2013}
\begin{equation}
|\Psi[Q,R]\rangle \equiv \tr\left(\mathcal{P}\exp\left[\int_{-\infty}^{\infty} Q\otimes \mathbb{I} +
R\otimes\hat{\psi}_x^\dag\, dx\right]\right)|\Omega\rangle,
\end{equation}
where $\mathcal{P}\exp$ denotes the \emph{path ordering} of the argument from left to right for increasing values
of $x$. The operators $Q$ and $R$ act on an auxiliary space $\mathbb{C}^D$, and $|\Omega\rangle$
is the Fock vacuum. The variational
parameters specifying the cMPS are precisely the two $D\times D$ matrices $Q$ and $R$.
The Lieb-Liniger hamiltonian $\hat{H}$ (in the presence of a chemical potential) is comprised of three terms
$\hat{H} = \hat{T} + \hat{W} + \hat{N}$, and the expectation values of these three terms can be readily computed
\cite{Haegeman2013} in terms of the variational parameters according to
\begin{align}
\langle \Psi[Q,R]|\hat{T}|\Psi[Q,R]\rangle &= \tr\left([Q,R]^\dag [Q,R] \rho_{\text{ss}}\right), \\
\langle \Psi[Q,R]|\hat{W}|\Psi[Q,R]\rangle &= v\tr\left({R^\dag}^2R^2 \rho_{\text{ss}}\right),
\quad\text{and} \\
\langle \Psi[Q,R]|\hat{T}|\Psi[Q,R]\rangle &= -\mu\tr\left(R^\dag R\rho_{\text{ss}}\right),
\end{align}
where $\rho_{\text{ss}}$ is the solution of the matrix equation
\begin{equation}
0 = -i[K,\rho] + R\rho R^\dag - \frac{1}{2}\{R^\dag R, \rho\},
\end{equation}
and $K = iQ + \frac{i}{2}R^\dag R$.
In order to find the variational minimum of
\begin{eqnarray}
E[v,\mu;Q,R] &\equiv& \langle
\Psi[Q,R]|\hat{H}(v,\mu)|\Psi[Q,R]\rangle
\nonumber
\\
&& \hspace{-14mm}= \tr\left(\left\{[Q,R]^\dag [Q,R] + v{R^\dag}^2R^2 -\mu R^\dag R
\right\} \rho_{\text{ss}}\right)
\nonumber
\\
\end{eqnarray}
 with respect to $Q$ and $R$ a \emph{tangent-plane} method using the
\emph{time-dependent variational principle} (TDVP) in imaginary time was exploited \cite{Haegeman2011}. This method proceeds as follows. Firstly, $v$ and $\mu$ are selected. Then a value
$D$ as large as possible is chosen. An initial guess $|\Psi[Q_0,R_0]\rangle$ for the ground state results from random choice of $Q_0$ and $R_0$. Also a tolerance $\eta$ and a step size $\delta$ is selected. Set $j=0$ and
perform the following sequence of operations until the desired convergence is reached.
\begin{enumerate}
\item Calculate the gradient $\nabla_{z} |\Psi[Q_j,R_j]\rangle$, where $z$ is a $2D^2$ vector containing the
entries of $Q_j$ and $R_j$ in, say, lexicographic order. Thus $\nabla_{z} |\Psi[Q_j,R_j]\rangle$ is a vector of
$2D^2$ cMPS states.
\item Calculate the gradient $\nabla_z E[Q_j,R_j]$ of the energy expecation value.
\item Calculate the inverse $G^{-1}$ of the {Gram matrix} $G_{z,z'} \equiv (\nabla_{z}
|\Psi[Q_j,R_j]\rangle)^\dag\nabla_{z'} |\Psi[Q_j,R_j]\rangle$.
\item Set $z_{j+1} = z_j - \delta \sum_{z'=1}^{2D^2}G^{-1}_{z,z'}\nabla_{z'} E[Q_j,R_j]$.
\item Set $j = j+1$, unpack $z_{j+1}$ into the two matrices $Q_{j+1}$ and $R_{j+1}$, and repeat step (1)
until convergence of the energy expectation values reaches the prespecified tolerance $\eta$.
\end{enumerate}
After the above algorithm has terminated the cMPS corresponding to unit density (and at the rescaled interaction
parameter) is obtained via the rescaling procedure described in the previous section.
This method was used to calculate a cMPS representation for the Lieb-Liniger ground state across a range of
interaction parameters from $v = 0$ to $v = 1000$. The value $D=14$ was used throughout as the results so
obtained are indistinguishable from the known exact solutions at $v=0,2,\infty$. For the other values of the
interaction parameter the accumulated errors were estimated and found to be negligible.
\bibliographystyle{apsrev4-1}
\end{document}